\documentclass[letterpaper,english,reprint,aps,preprintnumber]{revtex4-1}
\usepackage[T1]{fontenc}
\usepackage[utf8]{inputenc}
\usepackage{bm}
\usepackage{amsmath}
\usepackage{amssymb}

\makeatletter

\pdfpageheight\paperheight
\pdfpagewidth\paperwidth

\usepackage{babel}

\makeatother

\usepackage{babel}
\begin{document}

\preprint{KOBE-COSMO-17-14}

\title{Electromagnetic waves propagating in the string axiverse}

\author{Daiske Yoshida}
\email{dice-k.yoshida@stu.kobe-u.ac.jp}

\author{Jiro Soda}
\email{jiro@phys.sci.kobe-u.ac.jp}

\affiliation{Department of Physics, Kobe University, Kobe, 657-8501, Japan}

\date{\today}
\begin{abstract}
It is widely believed that axions are ubiquitous in string theory
and could be the dark matter. The peculiar features of the axion dark
matter are coherent oscillations and a coupling to the electromagnetic
field through the Chern-Simons term. In this paper, we study consequences
of these two features of the axion with the mass in a range from $10^{-13}\,{\rm eV}$
to $10^{3}\,{\rm eV}$. First, we study the parametric resonance of
electromagnetic waves induced by the coherent oscillation of the axion.
As a result of the resonance, the amplitude of the electromagnetic
waves is enhanced and the circularly polarized monochromatic waves
will be generated. Second, we study the velocity of light in the background
of the axion dark matter. In the presence of the Chern-Simons term,
the dispersion relation is modified and the speed of light will oscillate
in time. It turns out that the change of speed of light would be difficult
to observe. We argue that the future radio wave observations of the
resonance can give rise to a stronger constraint on the coupling constant
and/or the density of the axion dark matter. 
\end{abstract}
\maketitle

\section{Introduction}

According to string theory, axions are ubiquitous in the universe,
dubbed the string axiverse~\cite{Svrcek:2006yi,Arvanitaki:2009fg,Cicoli:2012sz,Halverson:2017deq,Stott:2017hvl}.
Remarkably, the axions could be a dark component of the universe and
might be a dominant piece of the dark matter~\cite{Preskill:1982cy,Abbott:1982af,Dine:1982ah,Hu:2000ke,Marsh:2015xka,Hui:2016ltb,Lee:2017qve}.
In fact, it is difficult to discriminate between the axion dark matter
and the cold dark matter on large scales. Therefore, it is important
to find a method for proving the existence of the axions.

The key feature of the axion dark matter is its coherent oscillation.
In particular, if the axion has the mass $10^{-23}\,{\rm eV}$, the
time scale of the oscillation is a few years and the oscillation produces
the oscillation in the gravitational potential. Hence, one can use
pulsar timing arrays to observe oscillating gravitational potential~\cite{Khmelnitsky:2013lxt,Porayko:2014rfa,Aoki:2016mtn,Aoki:2017ehb}.
There are other methods proposed for detecting the axion dark matter,
for example, the super-radiance instability of the axion field in
the rotating black holes constraining the mass range $10^{-20}\sim10^{-10}\,{\rm eV}$~\cite{Arvanitaki:2009fg,Arvanitaki:2010sy,Yoshino:2015nsa,Brito:2017zvb},
gravitational wave interferometers for probing the axion with mass
$10^{-22}\sim10^{-20}\,{\rm eV}$~\cite{Aoki:2016kwl}, the dynamical
resonance of the binary pulsars probing the mass range $10^{-23}\sim10^{-21}\,{\rm eV}$~\cite{Blas:2016ddr},
and cosmological axion oscillations for exploring a wide mass range~\cite{Soda:2017dsu,Zhang:2017dpp}.

Recently, we have studied the gravitational waves in dynamical Chern-Simons
gravity in the axion dark matter background~\cite{Yoshida:2017cjl}.
Then, we found that there occurs the parametric resonance of gravitational
waves with parity-violation, that is, circularly polarized gravitational
waves which allows us to probe the axions with the mass range $10^{-14}\sim10^{-10}\,{\rm eV}$.

Apparently, we can expect the same phenomena for electromagnetic waves.
Since electromagnetic waves are often used to explore the universe,
it is worth studying the phenomena in detail. The electrodynamics
in the presence of the axion is called the axion electrodynamics~\cite{Wilczek:1987mv}
which has the Chern-Simons coupling between the axion and the gauge
field. We see this interaction induces the parametric resonance of
electromagnetic waves and also yields to the oscillation of the speed
of light in time. In this paper, we study these two effects to gives
rise to a new way to explore the axion dark matter in a mass range
$10^{-13}\sim10^{3}\,{\rm eV}$ corresponding to the observable frequency
range of electromagnetic waves $10{\rm Hz}\sim10^{5}\,{\rm THz}$.
Note that the axions with the mass above $10^{3}$ eV are unstable
against decaying into photons~\cite{Lee:2017qve,Hu:2000ke,Marsh:2015xka,Hui:2016ltb}.

The organization of the paper is as follows. In Sec. II, we introduce
the axion electrodynamics. In Sec. III, we derive wave equations in
the oscillating axion background. In Sec. IV, we study the parametric
resonance in the axion background. In Sec. V, we investigate the speed
of light. The final section is devoted to conclusion.

\section{Axion electrodynamics}

The action of the axion electrodynamics is given by 
\begin{equation}
S=S_{{\rm EM}}+S_{{\rm \Phi}}+S_{{\rm int}}\ ,
\end{equation}
where each part of this action reads 
\begin{equation}
\begin{aligned}S_{{\rm EM}}\equiv & \int dx^{4}\sqrt{-g}\left(-\frac{1}{4}F_{\mu\nu}F^{\mu\nu}\right),\\
S_{{\rm \Phi}}\equiv & \int dx^{4}\sqrt{-g}\left(-\frac{1}{2}\left(\nabla_{\!\mu}{\rm \Phi}\right)\left(\nabla^{\mu}{\rm \Phi}\right)-U({\rm \Phi})\right)\ ,\\
S_{{\rm int}}\equiv & \int dx^{4}\sqrt{-g}\left(-\frac{\lambda}{4}{\rm \Phi}F_{\mu\nu}\tilde{F}^{\mu\nu}\right).
\end{aligned}
\end{equation}
Here $\lambda$ is a coupling constant, $U({\rm \Phi})$ is a potential
function for an axion field ${\rm \Phi}$, and $A^{\mu}=(A^{0},\bm{A})$
is a gauge field with the field strength $F_{\mu\nu}\equiv\nabla_{\!\mu}A_{\nu}-\nabla_{\!\nu}A_{\mu}$.
The dual of the field strength $\tilde{F}^{\mu\nu}$ is defined by
\begin{equation}
\tilde{F}^{\mu\nu}=\frac{1}{2}\epsilon^{\mu\nu\rho\sigma}F_{\rho\sigma},
\end{equation}
where the anti-symmetrical epsilon tensor $\epsilon^{\mu\nu\rho\sigma}$
is given by 
\begin{equation}
\epsilon^{\mu\nu\rho\sigma}\equiv\frac{1}{\sqrt{-g}}\tilde{\epsilon}^{\mu\nu\rho\sigma}\hspace{1em}\text{and}\hspace{1em}\tilde{\epsilon}^{0123}=+1.
\end{equation}
Here, $\tilde{\epsilon}^{\mu\nu\rho\sigma}$ is the Levi-Civita symbol.

From the above action, we get the equations of motion for the electromagnetic
waves

\begin{equation}
\nabla_{\!\mu}F^{\alpha\mu}+\frac{\lambda}{2}\epsilon^{\alpha\mu\nu\lambda}\left(\nabla_{\!\mu}{\rm \Phi}\right)F_{\nu\lambda}=0
\end{equation}
and the equation for the axion field 
\begin{equation}
\nabla_{\!\mu}\nabla^{\mu}{\rm \Phi}-\frac{d}{d{\rm \Phi}}U({\rm \Phi})=\frac{\lambda}{4}F_{\mu\nu}\tilde{F}^{\mu\nu}.
\end{equation}
Now, we can study electromagnetic wave propagation in the axion background.

\section{Wave equations in the axiverse }

We assume the background spacetime is the Minkowski spacetime, because
the dynamics of the cosmic expansion can be neglected on inter-galactic
scales~\cite{Yoshida:2017fao}. Then, the metric reads 
\begin{equation}
\begin{aligned}ds^{2} & =\,\eta_{\mu\nu}\,dx^{\mu}\,dx^{\nu}\\
 & =\,-dt^{2}+dx^{2}+dy^{2}+dz^{2}\ .
\end{aligned}
\end{equation}
Now, the covariant derivative is simply reduced to a partial derivative
$\partial_{\mu}$. We are interested in the time-evolution of the
gauge field in the axion background. The gauge field is considered
as the perturbed field $A_{\mu}=\delta A_{\mu}$. Next, we consider
a homogeneous axion background 
\begin{equation}
{\rm \Phi}(t,\bm{x})={\rm \Phi}(t)\ .
\end{equation}
Then, the equation of motion of axion is given by 
\begin{equation}
(\partial_{t}^{2}+m^{2}){\rm \Phi}(t)\simeq0\ .
\end{equation}
Here, we assumed the potential of the axion as 
\begin{equation}
U({\rm {\rm \Phi}})=\frac{1}{2}m^{2}{\rm \Phi}^{2}\ .
\end{equation}
It is easy to obtain the solution 
\begin{equation}
{\rm \Phi}(t)={\rm \Phi}_{0}\cos(m\,t)\ ,
\end{equation}
where ${\rm \Phi}_{0}$ is determined by the density of the dark matter
$\rho$ and the mass of the axion $m$ as 
\begin{equation}
{\rm \Phi}_{0}=\frac{\sqrt{2\rho}}{m}\ .
\end{equation}
The equations of motion of the axion electrodynamics can be deduced
as 
\begin{equation}
\begin{aligned}\partial_{\mu}\delta\!F^{0\mu} & =0\ ,\\
\partial_{\mu}\delta\!F^{i\mu}-\lambda\epsilon^{ijk}\left(\partial_{0}\Phi\right)\partial_{j}\delta\!A_{k} & =0.
\end{aligned}
\end{equation}
Here, the epsilon tensor in this coordinate system is defined as 
\begin{equation}
\epsilon^{ijk}\equiv\epsilon^{tijk}.
\end{equation}
The time-component of the modified Maxwell equation is the same as
the conventional Maxwell equation.

This modified Maxwell theory is invariant under the gauge transformation,
\begin{equation}
A_{\mu}\rightarrow A'_{\mu}=A_{\mu}+\partial_{\mu}\Lambda.
\end{equation}
So, we can adopt the radiation gauge for the electromagnetic field,
\begin{equation}
\delta\!A^{0}=0,\hspace{1em}\nabla\cdot\delta\!\bm{A}=0,
\end{equation}
and we get the wave equations of the axion electrodynamics, 
\begin{equation}
\Box\,\delta\!\bm{A}+\lambda\left(\partial_{0}\Phi\right)(\nabla\times\delta\!\bm{A})=0,
\end{equation}
where we defined the derivative operators $\Box\equiv\nabla_{\!\mu}\nabla^{\mu}$
and $\nabla\equiv(\partial_{x},\partial_{y},\partial_{z})$.

We can diagonalize the wave equations with the circular polarization
basis. In Fourier space, the vector field $\delta\!\bm{A}$ is expressed
by 
\begin{equation}
\delta\!\bm{A}\equiv\int\bm{a}(t)\,e^{i\bm{k}\cdot\bm{x}}d\bm{k}\ ,
\end{equation}
where $\bm{k}$ is the wave number vector. The transverse gauge condition
can be written as 
\begin{equation}
\bm{k}\cdot\bm{a}(t)=0\ .
\end{equation}
We can take polarization basis vectors, $\bm{e}_{(1)},\,\,\bm{e}_{(2)}$,
satisfying the following conditions 
\begin{eqnarray}
 &  & \bm{e}_{(I)}\cdot\bm{k}=0,\\
 &  & \bm{e}_{(I)}\cdot\bm{e}_{(J)}=\delta_{IJ}\ ,\hspace{1em}\text{for }I,\,J=(1,2)\\
 &  & \bm{e}_{(1)}\times\bm{e}_{(2)}=\frac{\bm{k}}{k}\ .
\end{eqnarray}
Here, we defined $k=|\bm{k}|$. Thus, the Fourier coefficient $\bm{a}(t)$
is expanded as 
\begin{equation}
\bm{a}(t)=\sum_{I=1,2}a_{I}(t)\bm{e}_{(I)}.
\end{equation}
Alternatively, we can use the circular polarization basis 
\begin{equation}
\bm{e}_{{\rm R}}\equiv\frac{\bm{e}_{(1)}+i\bm{e}_{(2)}}{\sqrt{2}}\hspace{1em}\text{and}\hspace{1em}\bm{e}_{{\rm L}}\equiv\frac{\bm{e}_{(1)}-i\bm{e}_{(2)}}{\sqrt{2}}.
\end{equation}
Now, the Fourier coefficient $\bm{a}(t)$ is expanded as 
\begin{equation}
\bm{a}(t)=\sum_{B={\rm L},{\rm R}}a_{B}(t)\bm{e}_{B}.
\end{equation}
Note that the components are related as 
\begin{equation}
a_{{\rm R}}=a_{(1)}-i\,a_{(2)},\hspace{1em}a_{{\rm L}}=a_{(1)}+i\,a_{(2)}.
\end{equation}
This basis is useful for studying the parity violation. Using the
relation 
\begin{eqnarray}
\epsilon^{ijk}\frac{k^{j}}{k}\bm{e}_{R/L}^{k}=\mp i\bm{e}_{R/L}^{i}
\end{eqnarray}
we can diagonalize the wave equations as 
\begin{equation}
{\displaystyle \ddot{a}_{B}+k^{2}\left(1+\epsilon_{B}\lambda\frac{m}{k}\Phi_{0}\sin(mt)\right)a_{B}=0}
\end{equation}
where 
\begin{equation}
\begin{array}{c}
\epsilon_{B}=\begin{cases}
1 & :B={\rm R}\ ,\\
-1 & :B={\rm L}\ .
\end{cases}\end{array}
\end{equation}
This equation is nothing but the Mathieu equation describing the parametric
resonance. Therefore, the growth rate is given by 
\begin{eqnarray}
\Gamma=\frac{1}{4}\lambda m\Phi_{0}\ .
\end{eqnarray}
Since the axion has a non-trivial profile, the parity symmetry is
violated in the equation of motion. Thus, the circular polarization
should be generated. To be more precise, it is useful to define the
polarization-rate of the electromagnetic field 
\begin{equation}
{\rm parity}(t)\equiv\frac{\left|\dot{a}_{{\rm R}}\right|^{2}-\left|\dot{a}_{{\rm L}}\right|^{2}}{\left|\dot{a}_{{\rm R}}\right|^{2}+\left|\dot{a}_{{\rm L}}\right|^{2}}.
\end{equation}
Due to the parametric amplification, the growth of one of the modes
is larger that the other mode. In that case, we should have ${\rm parity}(t)\simeq\pm1$.
Moreover, since the dispersion relation is modified by the axion,
the speed of light is oscillating. We study the effects of these phenomena
on electromagnetic waves in the following.

\section{Parametric resonance}

We assume that a lot of clumps whose sizes are about the Jeans length
$L_{{\rm a}}$ exist in the core of Galaxy and the axion is coherently
oscillating there. These fuzzy object have the interaction with the
electromagnetic fields through the Chern-Simons coupling. Thus, the
coherent oscillations of the axion induce the parametric resonance
of electromagnetic waves.

From the general theory of the parametric resonance, the resonance
wave number $k_{{\rm r}}$ is given by 
\begin{equation}
k_{{\rm r}}=\frac{m}{2}.
\end{equation}
It is convenient to convert $k_{{\rm r}}$ into the resonance frequency
$f_{{\rm r}}$ of the waves as 
\begin{equation}
f_{{\rm r}}=1.2\times10^{4}\,{\rm Hz}\times\left(\frac{m}{10^{-10}\,{\rm eV}}\right).
\end{equation}
This frequency corresponds to VLF (very low frequency) band, $3\sim30\,{\rm kHz}$.
The existing FAST (Five-hundred-meter Aperture Spherical radio Telescope)
has the frequency band from $70\,{\rm MHz}$ to $3\,{\rm GHz}$ in
\cite{Nan:2011um}. Hence, this detector can survey the mass range
from $10^{-7}\,{\rm eV}$ to $10^{-5}\,{\rm eV}$. The SKA (Square
Kilometre Array) has the frequency from $50$ MHz to $350$ MHz (SKA-low)
and from $350$ MHz to $14$ GHz~\cite{Dewdney2015}. Now, this detector
will survey the mass range from $10^{-7}\,{\rm eV}$ to $10^{-4}\,{\rm eV}$.
If we consider the heavier axion with mass $m\sim1\,{\rm eV}$, the
resonance frequency is that of the visible light around $10^{2}\,{\rm THz}$.

On halo scales of the Galaxy, the energy density of the axion dark
matter is about $0.3\,{\rm GeV/cm^{3}}$. Hence, the growth rate can
be estimated as 
\begin{equation}
\Gamma_{{\rm max}}=5.4\times10^{-29}\,{\rm eV}\times\left(\frac{\lambda}{(10^{16}\,{\rm GeV})^{-1}}\right)\sqrt{\frac{\rho}{0.3\,{\rm GeV/cm^{3}}}}.
\end{equation}
Notice that this quantity is independent of the mass of the axion.
In fact, the growth rate is determined by the coupling constant and
the energy density of the axion dark matter. From this growth rate,
we can estimate the time scale, $t_{\times10}$, for the amplitude
to become ten times, as 
\begin{equation}
t_{\times10}=4.3\times10^{28}\,{\rm eV}^{-1}\times\left(\frac{(10^{16}\,{\rm GeV})^{-1}}{\lambda}\right)\sqrt{\frac{0.3\,{\rm GeV/cm^{3}}}{\rho}}.
\end{equation}
Note that the time corresponding to 1pc is given by $t_{1{\rm pc}}\simeq1.6\times10^{23}\,{\rm eV^{-1}}$.
Thus, after the $10$ Mpc propagation, the amplitude will be enhanced
by $10^{10^{2}}$ times. Therefore, we can obtain a stringent constrain
on the coupling constant and/or the fraction of the axion dark matter
in the universe.

The parametric resonance occurs in the frequency band 
\begin{equation}
f_{{\rm r}}-\frac{\Delta f}{2}\lesssim f_{{\rm r}}\lesssim f_{{\rm r}}+\frac{\Delta f}{2}\ ,
\end{equation}
where $\Delta f$ is given by 
\begin{equation}
\Delta f=2.6\times10^{-14}\,{\rm Hz}\times\left(\frac{\lambda}{(10^{16}\,{\rm GeV})^{-1}}\right)\sqrt{\frac{\rho}{0.3\,{\rm GeV/cm^{3}}}}\ .
\end{equation}
Since the band is very narrow, the circularly polarized monochromatic
wave grows sharply at the resonance frequency.

If the electromagnetic waves go through near the core of the Galaxy,
the energy density of dark matter gets enhanced 
\begin{equation}
\rho\lesssim0.3\times10^{6}\,{\rm GeV/cm^{3}}.
\end{equation}
In this situation, $t_{\times10}$ becomes 
\begin{equation}
\begin{aligned}t_{\times10}= & 4.3\times10^{25}\,{\rm eV}^{-1}\\
 & \hspace{1em}\times\left(\frac{(10^{16}\,{\rm GeV})^{-1}}{\lambda}\right)\sqrt{\frac{0.3\times10^{6}\,{\rm GeV/cm^{3}}}{\rho}}
\end{aligned}
\end{equation}
From this estimation, the amplitude of waves going through the Galaxy
core is further amplified by about $10^{2}$ times. At the resonance
frequency, when the amplitudes of waves are highly amplified, the
electromagnetic wave should be fully polarized, namely, ${\rm parity}(t)\simeq\pm1$.

If we detected the resonance signal, we would argue that the axion
dark matter exist. If we did not detect the resonance signal, we would
be able to give the constraint on the energy density or the coupling
constant. Therefore, we can say that the future very long wavelength
radio wave observations of this effect can give rise to stronger constraints
on the coupling constant and/or the density of the axion dark matter.

\section{The speed of light}

In axion electrodynamics, the dispersion relation in the axion background
reads 
\begin{equation}
\omega^{2}=k^{2}\left(1+\epsilon_{A}\lambda\frac{m}{k}\Phi_{0}\,\sin\left(mt\right)\right).
\end{equation}
The phase velocity $v_{{\rm p}}$ is given by 
\[
v_{{\rm p}}\equiv\frac{\omega}{k}=\sqrt{1+\epsilon_{A}\lambda\frac{m}{k}\Phi_{0}\sin(mt)}.
\]
Then, the deviation from the speed of light $\delta c_{{\rm p}}$
is given by 
\begin{equation}
\begin{aligned}\delta c_{{\rm p}} & \equiv\left|v_{{\rm p}}-1\right|\\
 & \leq\left|\sqrt{1+\epsilon_{A}\lambda\frac{m}{k}\Phi_{0}}-1\right|\simeq\frac{\lambda\sqrt{\rho}}{\sqrt{2}k}.
\end{aligned}
\end{equation}
For example, if we observe the visible light which is in the wavelength
range $380\sim750\,{\rm nm}$, we find the relative deviation of the
speed of light: 
\begin{equation}
\begin{aligned}\delta c_{{\rm p}} & \simeq4.3\times10^{-29}\\
 & \times\left(\frac{\lambda}{(10^{16}\,{\rm GeV/cm^{3}})^{-1}}\right)\left(\frac{l_{{\rm em}}}{500\,{\rm nm}}\right)\sqrt{\frac{\rho}{0.3\,{\rm GeV/cm^{3}}}}.
\end{aligned}
\end{equation}
Here, $l_{{\rm em}}$ is the wave length of the visible light.

In fact, the group velocity is more relevant to observations. The
group velocity $v_{{\rm g}}$ is given by 
\begin{equation}
v_{{\rm g}}\equiv\frac{\partial\omega}{\partial k}=\frac{1}{2\omega}\left(2k+\epsilon_{A}\lambda m\Phi_{0}\sin\left(mt\right)\right).
\end{equation}
The deviation from the speed of light $\delta c_{{\rm g}}$ is given
by 
\[
\begin{aligned}\delta c_{{\rm g}} & \equiv\left|v_{{\rm g}}-1\right|\\
 & \simeq\left|1-\frac{1}{4}\lambda^{2}\frac{m^{2}}{k^{2}}\Phi_{0}^{2}\sin^{2}\left(mt\right)-1\right|\lesssim\frac{\lambda^{2}\rho}{2k^{2}}.
\end{aligned}
\]
Notice that the linear term is canceled out in the above formula~\footnote{We thank Tomohiro Fujita for pointing out this fact.}
and the deviation of the group velocity is given by the square of
that of the phase velocity 
\[
\delta c_{{\rm g}}=(\delta c_{{\rm p}})^{2}.
\]
Thus, we can estimate $\delta c_{{\rm g}}$ as 
\[
\begin{aligned}\delta c_{{\rm g}} & \simeq1.8\times10^{-49}\\
 & \times\left(\frac{\lambda}{(10^{16}\,{\rm GeV/cm^{3}})^{-1}}\right)^{2}\left(\frac{l_{{\rm em}}}{500\,{\rm nm}}\right)^{2}\left(\frac{\rho}{0.3\,{\rm GeV/cm^{3}}}\right).
\end{aligned}
\]
The relative deviation from the speed of light $\delta c$ is constrained
by observations of gamma-ray bursts \cite{Nemiroff:2012} as 
\begin{equation}
\delta c\lesssim10^{-21}\ .
\end{equation}
Since $\delta c_{g}$ is much smaller than the current observational
constraint, we can say that there is no constraint on the energy density
of axion field or the coupling constant from the speed of light.

\section{Conclusion}

Since the axion is one of the candidates for the dark matter, it is
worth seeking a method for detecting axion. In this paper, we considered
the axion with the mass range from $10^{-13}\,{\rm eV}$ to $10^{3}\,{\rm eV}$.
We focused on two consequences of the coherent oscillation of the
axion dark matter and a coupling to the electromagnetic field through
the Chern-Simons term. First, we studied the parametric resonance
of the gauge field induced by the coherently oscillating axion. It
turned out that, as a result of the resonance, the amplitude of the
electromagnetic waves is enhanced and the circularly polarized monochromatic
waves are generated. We found that the future very long wavelength
radio wave observations of this effect can give rise to stronger constraints
on the coupling constant and/or the density of the axion dark matter.
Second, we studied the velocity of light in the background of the
axion dark matter. We found that the dispersion relation is modified
and the speed of light shows oscillations in time, but this modification
is too tiny to be observed.

In this paper, we have discussed the modification of the dispersion
relations which leads to the change of the speed of light. However,
this effect was very small in axion electrodynamics. This would also
happen to gravitational waves. We report the detailed analysis in
a future work. 
\begin{acknowledgments}
We would like to thank Asuka Ito for useful discussions. We are grateful
to Tomohiro Fujita for useful comments. D.Y. was supported by Grant-in-Aid
for JSPS Research Fellow and JSPS KAKENHI Grant Number JP17J00490.
J.S. was in part supported by JSPS KAKENHI Grant Numbers JP17H02894,
JP17K18778, JP15H05895 and JP17H06359. 
\end{acknowledgments}

 \bibliographystyle{apsrev4-1}
\bibliography{AxionResonanceEM}

\end{document}